\documentstyle{article}
\begin{document}
\title{First International Conference on {\it Unconventional Models of
Computation UMC'98}\\An Unconventional Review}
\author{K. Svozil\\
 {\small Institut f\"ur Theoretische Physik},
  {\small Technische Universit\"at Wien }     \\
  {\small Wiedner Hauptstra\ss e 8-10/136},
  {\small A-1040 Vienna, Austria   }            \\
  {\small e-mail: svozil@tph.tuwien.ac.at},
  {\small www: http://tph.tuwien.ac.at/$\widetilde{\;}$svozil}}
\maketitle

\begin{flushright}
{\scriptsize
http://tph.tuwien.ac.at/$\widetilde{\;}$svozil/publ/umc98.tex}
\end{flushright}

Its January---summertime! The sun
is up high, the surf is good, the sky is blue and the sand is golden and
hot. As these lines are written, the author has settled on the beautiful
beaches of New Zealand's Northland.
Unconventional methods of computation seem all but a remote
possibility.
And yet, many of today's speculations become tomorrow's physics and
 key technologies of the future.

The question at stake: what kind of information processing technology
will be used by my two children, now playing sand, when they are of my
age? On an even more fundamental level, what will the term
``information'' mean to them?

Will there be a technology shift from present-day silicon based
integrated circuits processing classical information to, say, quantum
devices? Will
quantum bits be the fundamental atoms of information? And will
DNA-based computation be a progressive concept of computation?

These were the sort of issues which brought together the researchers
participating in the
{\it First International Conference on Unconventional Models of
Computation (UMC'98)}, organized by Auckland Universities' {\it Centre
for
Discrete Mathematics and Theoretical Computer Science
(CDMTCS)}\footnote{http://www.cs.auckland.ac.nz/CDMTCS/index.html}, in
cooperation with the
{\it Santa Fe Institute}\footnote{http://www.santafe.edu}, represented
by Cristian Calude and John Casti, respectively.

Most of the participants traveled halfway around the globe to the
meeting. All of them were received with a warm welcome and a splendid
organization by the conference chair Cristian Calude, as well as by the
entire staff of the Department of Computer Science of the
University of Auckland headed by Peter Gibbons.
One indication of this was the timely delivery of the conference
proceedings \cite{umc98} by Springer Singapore, which were distributed
already at the beginning of the conference.

Let us come back to the scientific subject matters.
I would like to group them into the following four major categories
\begin{description}
\item[$\bullet$]
quantum computation and information theory,
\item[$\bullet$]
DNA-based computation,
\item[$\bullet$]
reversible computation,
\item[$\bullet$]
unicorns and  miscellaneous topics,
\end{description}
which will also determine the following brief and very
personal review. For a more complete and authentic picture, the reader
is strongly encouraged to read the conference proceedings \cite{umc98}.

Right at the beginning, Artur Eckert gave a keynote review of the
present status of quantum computation theory, which was later extended
with an emphasis to technological and experimental aspects by Seth
Lloyd and Jeff Kimble. Boris Pavlov proposed a new quantum switching
device based on resonance scattering.

In another contribution to the conference,
Elena Calude and Marjo Lipponen took up an idea of Edward Moore to
simulate quantum complementarity by finite automata.

Quantum computation appears to be a major emerging field of basic and
applied research. Like DNA computation, the stakes are high but so are
the difficulties and challenges.
In its present state, quantum computing mostly attempts to make use of
quantum
parallelism. In classical terms, a quantum computation may
pursue ``a very large number'' of sometimes contradictory tasks at a
single
computation step. This is due to the fact that, stated pointedly,
because what physicists
call ``coherent superposition'' of (classically contradictory) {\it
yes-no} bit states, quantum information may represent both classically
mutually exclusive bit states at once. These can then be co-processed
simultaneously by applying reversible one-to-one unitary
transformations. In this sense, any classical system in which $n$
classical
bit values are stored and processed
is outperformed by a quantum bit system by an exponential factor $2^n$,
containing all the $2^n$ classical bit values at once.

As can be expected, the exponential speedup comes with a price tag.
There is the so-called problem of
decoherence. Due to what is conventionally referred to as the wave
function collapse by the interaction of quantum bits with their
classically modeled environment, they decay and degenerate
uncontrollably into classical bits. But there seem to be least a partial
remedy. Very clever error correction strategies attempt to protect the
quantum bits. They too may have their price in terms of  an
increased slowdown. Research in this area is fascinating and far from
being completed.

Due to quantum complementarity, the quantum bits themselves are not
completely
measurable. Therefore, any meaningful result needs to be suitably
encoded into interference patterns. For the same reason, Shor's
prominent quantum algorithm \cite{shor:94} for factoring in quantum
polynomial time is probabilistic.

It may be suspected \cite{gottlob}
that,  due to the necessity of reaching reasonable fringe contrast
in the interference patterns, such methods can only be effectively
applied to the class of problems
UP having a unique solution; or to problems having ``close to unique''
solutions insofar as the number of actual solutions increases ``much
slower''
(e.g., polynomial) than the possible solutions.
Personally, I  confess that, having been deeply impressed by the
so called ``interaction-free schemes of measurement''
\cite{elitzur-vaidman:1,kwhz:95,benn:94}, I am more inclined to consider
``free lunch scenarios'' in quantum information and computation theory
as before.

Both Cristian Calude and Artur Eckert mentioned possible consequences
for proof theory. Additional attention, I believe,
should be given  to a revision
of recursive function theory due to the possibility to represent
classically contradicting bit states by a quantum bit which is in a
coherent
superposition thereof. Any diagonalization argument, for example,
includes a {\it not}-operation which results in the switch of one
classical bit state into the other. Quantum mechanically, this operation
has a fixed point quantum bit state which is a fifty-fifty
coherent mixture of the two classical states. Hence, for instance, the
proof of the recursive undecidability of the halting problem collapses,
because the {\it reductio ad absurdum} does no longer work---the
solution is
a fixed point quantum bit state; equivalent to the flipping of a
fair coin.


DNA-based computation has been developed extensively in the
contributions of
Gordon Alford,
Martin Amos,
Alan Gibbons,
Valeria Mihalache,
Gheorge P\u{a}un,
Animesh Ray,
 John Reif, and
Arto Salomaa, both from the theoretical and from the experimental side.
Again, research in that area is in full progress.

The idea here is to encode a problem into strands of DNA, profiting
again
from heavy parallelizing. This parallelizing might be limited by the
population size conceivable.


Another big topic of the conference was reversibility.
Reversible computation is characterized by one-to-one operations, by a
reversible, bijective evolution map of the computer states onto
themselves.
If only a finite number of such states are involved, this
amounts to their permutation.

In such a scheme, not a single bit
gets lost, and any piece of information (including the trash) remains in
the
computer forever. That may be good news for the case of decay and loss
of information, but it is bad news with respect to waste management.
There is no way of trashing garbage bits other than cleverly
compressing them and pile them ``high and deep.''
Stated differently:
in this restricted regime,
many-to-one operations such as deletion of bits are
not allowed.

Furthermore,
one-to-many operations such as copying are forbidden as well.
From this point of view, it may appear appropriate to develop
algorithmic
information theory in terms of prefix
reversible algorithms.

Classical continuum mechanics and electrodynamics are reversible
``at heart;'' that means that all  equations of motion are invariant
with
respect to reversing the arrow of time. Also quantum
computation is based upon unitary evolution of the quantum bit states
and thus is reversible. The
no-copy feature of reversible computation is for instance reflected by
the no-cloning theorem of quantum theory.

The MIT group headed by Thomas Knight presented silicon prototypes of
reversible computers based
on Fredkin's billiard ball model of universal computation, as well as
reversible gate
technology and memory management. The potential technological advantage
of reversible computers over irreversible ones lies in the fact that
reversible computation is not necessarily associated with energy
consumption and heat dissipation while the latter one is.
And since heat dissipation per computation step can be kept at arbitrary
low levels, when ``scaled up to very large sizes'' reversible
computation outperforms irreversible computation. Moreover, after all,
physics at very small scales {\it is} reversible.

We know since Landauer \cite{landauer:61} and Bennett \cite{bennett-82}
that any computation can be embedded into a reversible one. The trick is
to provide markers in order to make back-tracking possible, which amounts
to
memorizing the past states of the system. If no copying is allowed, this
may amount to large space overheads as compared to irreversible
computations. In
one contribution to the conference, Peter Hertling showed that there in
general  exist no better, more clever way to embed an irreversible
Cellular Automaton into reversible ones than doing just that.

In still another variation of the reversibility topic,
the author of this article constructed reversible finite automata via
permutation matrices.

There is one immediate algorithmic aspect of reversibility.
If reversible computation is just a
rephrasing, a permutation of the input, then what use it anyway?
The ``garbage in-garbage out'' metaphor is particularly pressing here.

Let us consider an example. Why should the product of the prime factors
$3\cdot 5$, obtained for instance by Shor's algorithm, be more
``informative'' than the number
15? And what
is the meaning of the term ``informative'' here? This issue seems to be
somewhat related to an old question in proof theory, in which sense a
proof of a statement is ``better'' then the mere knowledge that this
statement is true.
It seems that the term
``informative'' can only be given a subjective, idealistic meaning
devoid of any formal rigor. Because the number 15 would just be a
trivial rephrasing of its product of prime factors $3\cdot 5$, very much
so as its binary expansion 1111. That might eventually be bad news for
standard encryption techniques based on prime factorization. (Of course,
there might be ``a lot of '' reversible computation steps
necessary for this kind of ``trivial rephrasing,'' making
the equivalence ``very hard'' to verify.)


Christopher Moore gave a very interesting overview of analogue
computation. There he stressed the importance of perceiving the process
of computation as a {\em dynamical} one. In this sense, results from the
theory of dynamical systems can be applied to the theory of computation
and {\it vice versa}.
It seems to remain an interesting question if analogue computation,
which
in principle could make use of the infinite complexity of the objects
involved, could outperform Turing machines.

Herbert Wiklicky gave a very interesting framework for practically all
previously
discussed models of computation in terms of linearized systems or
Hilbert machines.

Speaking about unicorns in the little zoo of unconventional computers,
Zeno's argument against motion pops up here and there. In the context of
computations it appears as a model of computation using a Turing machine
with a geometrical increase of execution speed. In this way,
tasks which would need infinite computation time otherwise could be
completed in finite times.

Let me finish with a speculation.
Assume a pessimistic scenario in which we have convinced ourselves that
we cannot make use of the speedups promised by novel models of
computation. Then it might not be unreasonable to assume
that these alleged speedups emerge from an improper representation
of the natural phenomena, from our incorrect theoretical perception of
physical systems rather than from the phenomena themselves. That, of
course, would have far-reaching consequences for physics.


 \end{document}